# Improving the capacity of quantum dense coding by weak measurement and reversal measurement

Mao-Bin Tian(田茂彬), Guo-Feng Zhang(张国锋)[*]

*Key Laboratory of Micro-Nano Measurement-Manipulation and Physics (Ministry of Education), School of Physics and Nuclear Energy Engineering, Beihang University, Xueyuan Road No. 37, Beijing 100191, China*

**Abstract:** A protocol of quantum dense coding protection of two qubits is proposed in amplitude damping (AD) channel using weak measurement and reversal measurement. It is found that the capacity of quantum dense coding under the weak measurement and reversal measurement is always greater than that without weak measurement and reversal measurement. When the protocol is applied, for the AD channels with different damping coefficient, the result reflects that quantum entanglement can be protected and quantum dense coding becomes successful.

I. Introduction

In recent years, with the development of quantum mechanics and the establishment of quantum information science, quantum entanglement becomes a new research hotspot of theoretical physics. Quantum entanglement, which is a key part of quantum information processing[1], including quantum teleportation[2-3], quantum computing[4], quantum cryptography[5], quantum dense coding[6] and quantum secret sharing[7]. However, in realistic quantum information processing, entanglement is inevitably affected by the interaction between the system and environment, which leads to degradation. Furthermore, entanglement sudden death (ESD)[8–11] will appear in this case. Thus, it is very important to protect entanglement in quantum information processing.

---

[*] Correspondence and requests for materials should be addressed to G.Z.(gf1978zhang@buaa.edu.cn)

Weak measurement is a kind of partial collapse measurement, which was based on von Neumann measurement and positive operator valued measurement (POVM). We know that weak measurement is limited by the information extracted from the quantum system and can effectively prevent the quantum state of the measurement system from collapsing to its eigenstate randomly. Thus, quantum state can be reversed to its initial state with a certain probability under proper operation[12]. Since the seminal work by Aharonov, Albert, and Vaidman[13], weak measurement and reversal measurement have been studied in theory[14, 15] and realized in experiment[16]. For example, Liao[17] preserved entanglement and the fidelity of three-qubit quantum states by using quantum reversal operation. Environment-assisted quantum state restoration via weak measurements was argued about by Wang[18]. Sun[19] showed that entanglement protection of low-dimensional quantum systems by the means of weak measurement. Kim[20] demonstrated that weak measurement can effectively protect the entanglement of two qubits in the amplitude damped (AD) channel. In future, weak measurement will have important applications in the feedback control of quantum systems, observation of the spin hall effect of photons, protection of encoded arbitrary states and so on.

Quantum dense coding is an application of entanglement[21, 22] in quantum information processing. For the process of quantum dense coding, if Alice (the sender) wants to send more than one qubit of information, she only needs to operate a particle in the maximum entangled state. This result can be interpreted as a particle in two-qubit entangled state can transmit two bits of classical information and the capacity of source coding can be increased by quantum dense coding exponentially. Higher–dimensional entanglement provides more capacity of quantum dense coding over conventional qubit entanglement[23]. The capacity of quantum dense coding can be measured by the Holevo quantity:

$$\chi = S(\bar{\rho}) - S(\rho), \quad (1)$$

here, $\rho$ is the state shared between Alice and Bob, $\bar{\rho}$ represents the density matrix after quantum dense coding. $S$ is the von Neumann entropy and $\chi$ is the capacity of quantum dense coding. For the density matrix $\rho$, the von Neumann entropy is written explicitly:

$$S(\rho) = -\sum_Z \lambda_Z \log_2 \lambda_Z, \quad (2)$$

where the $\lambda_Z$ are eigenvalues of the density matrix $\rho$.

In this paper, quantum dense coding using weak measurement and reversal measurement in AD channel will be investigated in detail. We introduce two plans, as illustrated in FIG 1. The first plan is the initial state directly passage through the AD channel and then we make a dense coding. As an improvement on the former, the second plan is "weak measurement + AD channel + reversal measurement + dense coding". In this case, the capacity of quantum dense coding is better than that without weak measurement and reversal measurement. Moreover, it is found that the von Neumann entropy under weak measurement and reversal measurement is independent of the damping coefficient $d$ and can arrive at the maximum value by adjusting weak measurement strength $p$ and reversal measurement strength $q$.

The paper is structured as follows. Amplitude damped channel is introduced in Sec. II. And Sec. III is devoted to introducing weak measurement and quantum dense coding. The influences of damping coefficient $d$, weak measurement strength $p$ and reversal measurement strength $q$ on quantum dense coding are analyzed in Sec. IV. The conclusions are presented in Sec. V.

**II.  Amplitude damped channel**

There are three typical attenuation of single qubit in quantum information processing, but the amplitude damped (AD) noise model is classic between system and environment. For instance, the AD noise model can be used to describe the spontaneous emission of a photon by a two-level atom at low or zero temperature[2, 15].

For qubits, there are two configurations of 2-level system to be taken into account. The lower level and the upper level are respectively denoted as $|0\rangle$ and $|1\rangle$. If the environment is in a vacuum state, the AD noise of system and environment which corresponds to the spontaneous emission can be described as

$$U_{AE} = \begin{cases} |0\rangle_A|0\rangle_E \rightarrow |0\rangle_A|0\rangle_E \\ |1\rangle_A|0\rangle_E \rightarrow \sqrt{1-d}|1\rangle_A|0\rangle_E + \sqrt{d}|0\rangle_A|1\rangle_E \end{cases}, \quad (3)$$

where $d$ is the damping coefficient of AD channel with $d \in [0,1]$.

The initial state of the system can be defined as

$$|\psi\rangle_{AB} = \frac{1}{\sqrt{2}}(|00\rangle_{AB} + |11\rangle_{AB}), \quad (4)$$

after qubits transferring through the AD channel, the $AB$+environment combined system will evolve as follows

$$|X\rangle_{AB,E_1E_2} = \frac{1}{\sqrt{2}}\Big[|00\rangle_{AB}|00\rangle_{E_1E_2} + (1-d)|11\rangle_{AB}|00\rangle_{E_1E_2} + \sqrt{d(1-d)}|10\rangle_{AB}|01\rangle_{E_1E_2} + $$

$$\sqrt{d(1-d)}|01\rangle_{AB}|10\rangle_{E_1E_2} + d|00\rangle_{AB}|11\rangle_{E_1E_2}\Big], \quad (5)$$

here, $E_1$ and $E_2$ are the environment of particle $A$ and $B$, respectively.

By tracing over the freedoms of environments, the density matrix is given by

$$\rho_{AB} = \begin{pmatrix} \frac{1+d^2}{2} & 0 & 0 & \frac{1-d}{2} \\ 0 & \frac{(1-d)d}{2} & 0 & 0 \\ 0 & 0 & \frac{(1-d)d}{2} & 0 \\ \frac{1-d}{2} & 0 & 0 & \frac{(1-d)^2}{2} \end{pmatrix}. \quad (6)$$

The AD channel can also be measured for two qubits using the following Kraus operators

$$\rho(t) = \sum_{i,j=1}^{2}(\varepsilon_i^A \otimes \varepsilon_j^B)\rho(0)(\varepsilon_i^A \otimes \varepsilon_j^B)^\dagger, \tag{7}$$

where density matrixes of $\rho(0)$ and $\rho(t)$ stand for initial moment and arbitrary moments respectively. $\varepsilon_i$ ($i = 1,2$) are Kraus operators which can be written as

$$\varepsilon_1 = \begin{pmatrix} 1 & 0 \\ 0 & \sqrt{1-d} \end{pmatrix}, \quad \varepsilon_2 = \begin{pmatrix} 0 & \sqrt{d} \\ 0 & 0 \end{pmatrix}. \tag{8}$$

Mathematically, Kraus operators satisfy the completely positive and trace preserving relation.

$$\sum_{i,j=1}^{2}(\varepsilon_i \otimes \varepsilon_j)^\dagger(\varepsilon_i \otimes \varepsilon_j) = I. \tag{9}$$

### III. Weak measurement and quantum dense coding

In this paper, weak measurement operator and reversal measurement operator for two qubits can be written as:

$$M_w(p_1, p_2) = \begin{pmatrix} 1 & 0 \\ 0 & \sqrt{1-p_1} \end{pmatrix} \otimes \begin{pmatrix} 1 & 0 \\ 0 & \sqrt{1-p_2} \end{pmatrix},$$

$$M_{rev}(q_1, q_2) = \begin{pmatrix} \sqrt{1-q_1} & 0 \\ 0 & 1 \end{pmatrix} \otimes \begin{pmatrix} \sqrt{1-q_2} & 0 \\ 0 & 1 \end{pmatrix}, \tag{10}$$

$M_w(p_1, p_2)$ is a weak measurement operator, $p_1$ and $p_2$ are weak measurement strengths. In the same presentation, $M_{rev}(q_1, q_2)$ is the reversal measurement operator, $q_1$ and $q_2$ are reversal measurement strengths. For simplicity, we assumed that $p_1 = p_2 = p$ and $q_1 = q_2 = q$.

And then we make dense coding using entangled states of two-qubit system as a channel[24]. The set of mutually orthogonal unitary transformations of dense coding for two qubits are

$$U_{00}|x\rangle = |x\rangle,$$

$$U_{10}|x\rangle = e^{i\pi x}|x\rangle, \tag{11}$$

$$U_{01}|x\rangle = |x + 1(\mathrm{mod}2)\rangle,$$

$$U_{11}|x\rangle = e^{i\pi x}|x + 1(\mathrm{mod}2)\rangle,$$

where $|x\rangle$ is the single qubit computational basis. The average state of the ensemble of signal states generated by above equation is [25]

$$\rho^* = \frac{1}{4}\sum_0^3 (U_i \otimes I)\rho(U_i^\dagger \otimes I), \tag{12}$$

here, $I$ is the identity matrix of second order and $\rho$ represents the density matrix that needs to make dense coding. For simplicity, we suppose $0 \to 00$; $1 \to 01$; $2 \to 10$; $3 \to 11$.

### IV. Quantum dense coding in the amplitude damped channel

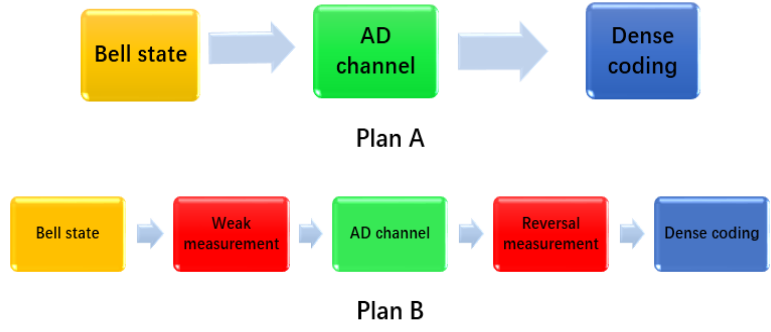

FIG.1: Plans for improving the capacity of quantum dense coding in AD channel. Plan A: Bell state goes through independent AD channel and then make dense coding. Plan B: it is similar to Plan A, but we make weak measurement before Bell states undergoes decoherence and reversal measurement after decoherence in AD channel.

The capacity of quantum dense coding as we mentioned above for different scheme

$$\chi_1 = S(\rho_1^*) - S(\rho_1),$$

$$\chi_2 = S(\rho_2^*) - S(\rho_2). \qquad (13)$$

Finally, we compare $\chi_1$ and $\chi_2$. If $\chi_2$ is bigger than $\chi_1$, it means that the capacity of quantum dense coding with weak measurement and reversal measurement is better than that without weak measurement and reversal measurement.

### A. Quantum dense coding without weak measurement and reversal measurement

After decoherence of the initial state in AD channel, the density matrix can be written as

$$\rho_1 = \begin{pmatrix} \frac{1+d^2}{2} & 0 & 0 & \frac{1-d}{2} \\ 0 & \frac{(1-d)d}{2} & 0 & 0 \\ 0 & 0 & \frac{(1-d)d}{2} & 0 \\ \frac{1-d}{2} & 0 & 0 & \frac{(1-d)^2}{2} \end{pmatrix}, \qquad (14)$$

here $\rho_1$ represents the density matrix without weak measurement and reversal measurement. According to Eq. (12), we can obtain the density matrix of $\rho_1^*$. Based on Eq. (13), the capacity of quantum dense coding can be obtained

$$\chi_1 = \frac{1-d}{2}\log_2\left(\frac{1-d}{4}\right) - \frac{1+d}{2}\log_2\left(\frac{1+d}{4}\right) + (1-d)d * \log_2\left[-\frac{1}{2}(-1+d)d\right]$$

$$+ \frac{1}{2}\left(1 - d + d^2 - \sqrt{1 - 2d + 2d^2}\right)\log_2\left[\frac{1}{2}\left(1 - d + d^2 - \sqrt{1 - 2d + 2d^2}\right)\right]$$

$$+ \frac{1}{2}\left(1 - d + d^2 + \sqrt{1 - 2d + 2d^2}\right)\log_2\left[\frac{1}{2}\left(1 - d + d^2 + \sqrt{1 - 2d + 2d^2}\right)\right]. \qquad (15)$$

Therefore, the relationship of the damping coefficient $d$ and the capacity of quantum dense coding without weak measurement and reversal measurement $\chi_1$ is plotted in Fig. 2:

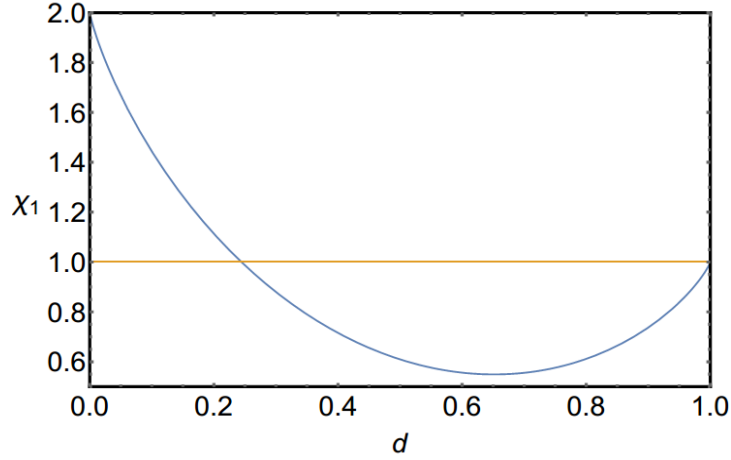

FIG.2: (Color online) The blue line represents the dependence of $\chi_1$ (the capacity of quantum dense coding without weak measurement and reversal measurement) on the damping coefficient $d$. The yellow line represents the capacity of quantum dense coding is equal to 1.

In quantum dense coding, if the capacity is lower than 1, it means quantum dense coding is not successful. From Fig. 2, it can be seen that the capacity of quantum dense coding without weak measurement and reversal measurement $\chi_1$ gets a minimum value 0.55 when the damping coefficient $d$ equals 0.652. Then the capacity of quantum dense coding is increased with increasing value of the damping coefficient and maintains a constant value of $\chi_1 = 1$ at $d = 1$. According to the curve in Fig. 2, it is clear that the capacity of quantum dense coding depends on the damping coefficient. If $d > 0.245$, quantum dense coding is not successful and we need to improve it. Therefore, we can draw a conclusion that quantum dense coding is not successful in AD channel for most situations.

**B. Quantum dense coding with weak measurement and reversal measurement**

After the sequence of weak measurement, decoherence, and reversal measurement, the density matrix can be described by

$$\rho_2 = \frac{1}{T}\begin{pmatrix} \rho_{11} & 0 & 0 & \rho_{14} \\ 0 & \rho_{22} & 0 & 0 \\ 0 & 0 & \rho_{33} & 0 \\ \rho_{41} & 0 & 0 & \rho_{44} \end{pmatrix}, \quad (16)$$

where $T$ is the normalized factor. In this equation:

$$\rho_{11} = \left[\frac{1}{2} + \frac{1}{2}d^2(1-p)^2\right](1-q)^2,$$

$$\rho_{22} = \rho_{33} = \frac{1}{2}(1-d)d(1-p)^2(1-q), \quad (17)$$

$$\rho_{44} = \frac{1}{2}(1-d)^2(1-q)^2,$$

$$\rho_{14} = \rho_{41} = \frac{1}{2}(1-d)(1-p)(1-q).$$

According to Eq. (2), the von Neumann entropy of $S(\rho_2)$ is shown in Fig.3:

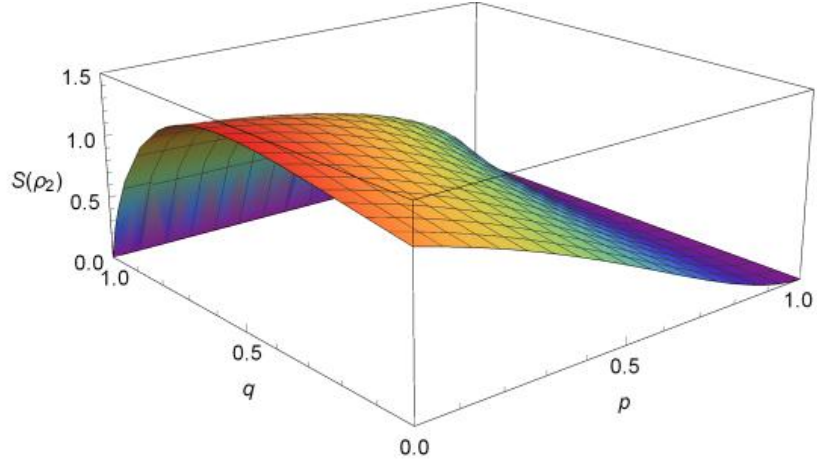

FIG.3: (Color online) The von Neumann entropy of $S(\rho_2)$ is plotted as functions of weak measurement strength $p$ and reversal measurement strength $q$ when $d = 0.5$.

In this case we can find that the von Neumann entropy decreases with the increase of weak measurement strength and reversal measurement strength. The von Neumann entropy will be zero if weak measurement strength and reversal measurement strength can get maximum value.

As for the research of $S(\rho_2^*)$, a new thinking can be put forward. Through calculations, we learn that no matter how the damping coefficient changes that maximum value can be got by adjusting weak measurement strength and reversal measurement strength. For the purpose of simplicity, we assume that the damping coefficient is a certainty and then discuss the effects of weak measurement and reversal measurement. And FIG.4 shows corresponding values for different weak measurement strength and reversal measurement strength. Every point in FIG.4 is the maximum value of the von Neumann entropy which is equal to 2. In our discussion, $d = 0.5$, $\chi_2$ must bigger than $\chi_1$ and 1 when weak measurement strength and reversal measurement strength are equal to 0.9. To make it easier to understand, we use a point in FIG.4 to analyze. For example, the von Neumann entropy $S(\rho_2^*) = 2$ for $p = 0.9$ and $q \approx 0.95$ when $d = 0.5$. From Fig. 3, it can be calculated that $S(\rho_2) \approx 0.33$ for the same damping coefficient. According to Eq.(13) and Eq.(15), we can get $\chi_2 = S(\rho_2^*) - S(\rho_2) = 1.67$ and $\chi_1 = 0.61$. In this case, the capacity of quantum dense coding under the weak measurement and reversal measurement is greater than that without weak measurement and reversal measurement. Furthermore, the method of weak measurement and reversal measurement can be used for different damping coefficient, which can make dense coding successful and improve the capacity of dense coding in AD channel.

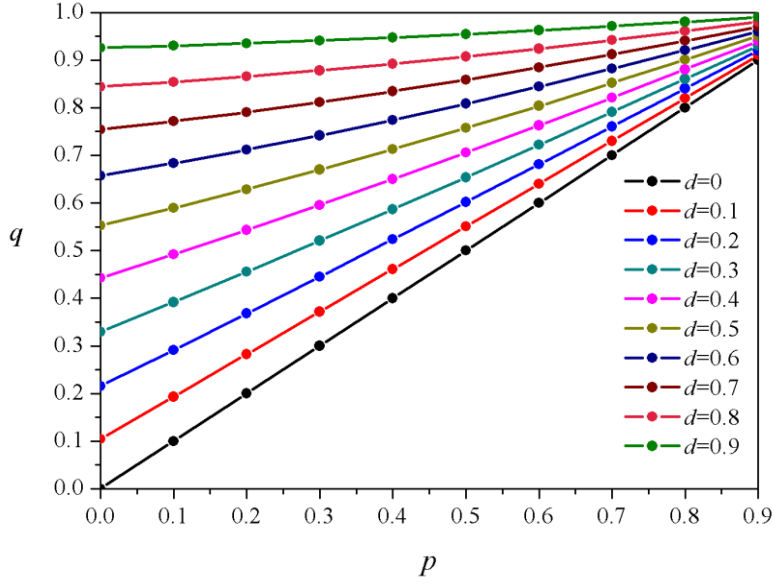

FIG.4: The maximum value of the von Neumann entropy $S(\rho_2^*)$ is plotted as functions of weak measurement strength $p$, reversal measurement strength $q$ and damping coefficient $d$.

For $d > 0.9$, the maximum value can be obtained. Because of the accuracy problem, we will not draw it here. In exceptional case, regardless of damping coefficient, weak measurement strength or reversal measurement strength is equal to 1, the finial result is $S(\rho_2) = 0$ and $S(\rho_2^*) = 1$.

The physical mechanism behind this is that a prior weak measurement can move the system close to ground state, which is not affected in AD channel. The decoherence is naturally suppressed in this "lethargic" state, and the entanglement is therefore preserved[26]. Even though the method decreases the probability of success, this procedure for dense coding is still useful.

## V. Conclusions

In this paper, we study a method to improve the capacity of quantum dense coding, and it can be used to decrease decoherence and protect entanglement in AD channel. Moreover, no matter how the damping coefficient changes that the maximum value of von Neumann entropy can be obtained by adjusting weak measurement strength and reversal measurement strength. By applying "weak measurement + AD channel + reversal measurement + dense coding" scheme, the decoherence can be efficiently suppressed and the capacity of quantum dense coding can be efficiently improved for the amplitude damping channel.

**Acknowledgements**

This work is supported by the National Natural Science Foundation of China (Grant No. 11574022).